\documentclass[a4paper,11pt]{article}
\usepackage{jinstpub} 
\usepackage{lineno}
\usepackage{siunitx}
\usepackage{graphicx}
\usepackage{adjustbox}

\title{\boldmath An ultra-light helium cooled pixel detector for the Mu3e experiment }

\author[a]{Thomas Theodor Rudzki}
\author[a]{Heiko Augustin}
\author[a]{David Maximilian Immig}
\author[a]{Ruben Kolb}
\author[a]{Lukas Mandok}
\collaboration[c]{on behalf of the Mu3e collaboration}
\affiliation[a]{Physikalisches Insitut,\\
Im Neuenheimer Feld 226, Heidelberg, Germany}

\emailAdd{rudzki@physi.uni-heidelberg.de}

\abstract{The Mu3e experiment searches for the lepton flavour violating decay $\mu^+ \rightarrow e^+ e^- e^+$ with an ultimate aimed sensitivity of $1$ event in $10^{16}$ decays.
To achieve this goal, the experiment must minimize the material budget per tracking layer to $X/X_0\approx \qty{0.1}{\percent}$ and use gaseous helium as coolant.
The pixel detector uses High-Voltage Monolithic Active Pixel Sensors (HV-MAPS) which are thinned down to \qty{50}{\micro\meter}.
Both helium cooling and HV-MAPS are a novelty for particle detectors.

Here, the work on successfully cooling a pixel tracker using gaseous helium is presented.
The thermal studies focus on the two inner tracking layers, the Mu3e vertex detector, and the first operation of a functional thin pixel detector cooled with gaseous helium.
The approach, which circulates gaseous helium under ambient pressure conditions with a gas temperature around \qty{0}{\degreeCelsius} using a miniature turbo compressor with a mass flow of \qty{2}{\gram/\second} allows the vertex detector to operate below \qty{70}{\degreeCelsius} at heat densities of up to \qty{350}{\milli\watt/\centi\meter^2}.
Finally, performance data of the final HV-MAPS used by Mu3e, the \textsc{MuPix11}, is presented.
These results demonstrate the feasibility of using HV-MAPS combined with gaseous helium as a coolant for an ultra-thin pixel detector exploring new frontiers in lepton flavor.}


\begin{document}
\maketitle
\flushbottom

\section{Introduction}
\label{sec:intro}

\begin{figure}[h]
    \centering
    \includegraphics[width=0.8\textwidth]{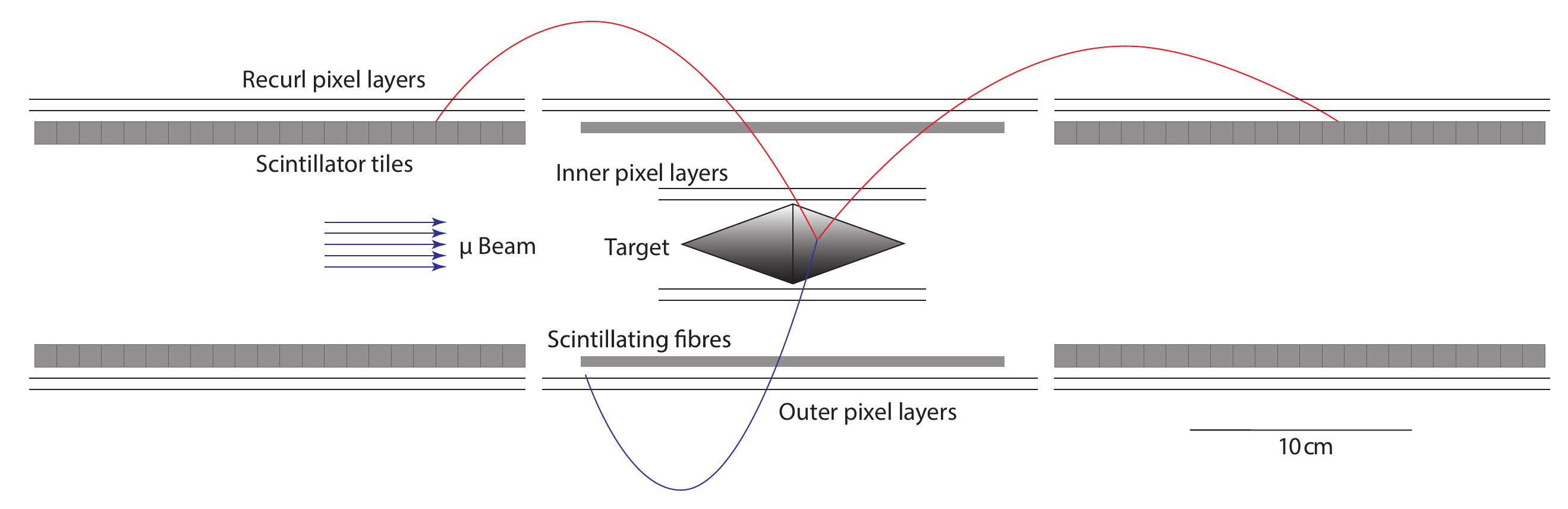}
    \caption{Schematic view on the Mu3e experiment. Muons are decaying at rest on the target, the decay products are detected by four central pixel layers. The particle momenta are determined by tracking their recurling trajectories in the magnetic field. Timing information is given by scintillators.}
    \label{fig:mu3e_schematic}
\end{figure}

Searching for the charged lepton flavour violating decay $\mu^+ \rightarrow e^+ e^- e^+$, the Mu3e experiment~\cite{mu3e} relies on an ultra-thin tracking detector.
The final single event sensitivity of 1 in $10^{16}$ decays is mainly limited by multiple-Coulomb scattering as described in~\cite{tdr}.
Thus, the design of the experiment focuses on radically minimizing the material budget in the active region.
High-voltage monolithic pixel sensors (HV-MAPS)~\cite{hvmaps} which are thinned down to \qty{50}{\micro\meter} were developed for this experiment.
All electrical connections and the mechanical support is provided by thin aluminium-polyimide high-density interconnects.
Together, the resulting material budget per tracking layer is \qty{0.115}{\percent\ X_0}.
The material minimization demands gaseous cooling of the pixel chips similar to STAR PXL~\cite{star}.
The decay products emerging from the target are detected by four tracking layers surrounding the target as depicted in \autoref{fig:mu3e_schematic}.
Inside the \qty{1}{\tesla} solenoidal magnetic field, the particles curl back and hit the outer pixel layers again.
The trajectory lengths of up to \qty{1}{\meter} would result in a material budget of \qty{0.33}{\percent\ X_0} for air as the medium.
To reduce scattering, gaseous helium ($\widehat{=}$~\qty{0.018}{\percent\ X_0 /\meter}) is chosen as coolant/medium instead, bringing the material budget to an unprecedented minimum.
The helium is circulated by novel miniature turbo compressors with high mass flows of up to \qty{16}{\gram/\second} while keeping the compression ratio below $\Pi < 1.1$ 

This paper summarizes in the following thermal studies of the gaseous helium cooling, as well as the first performance studies of the final HV-MAPS, the \textsc{MuPix11}.

\section{Gaseous helium cooling}

\begin{figure}[htbp]
\centering
\includegraphics[width=.53\textwidth]{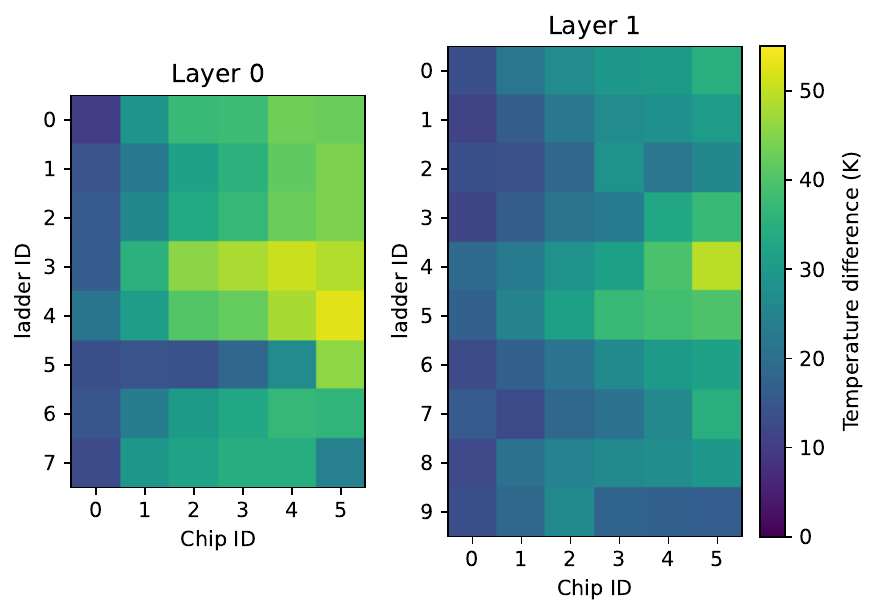}
\qquad
\adjustbox{trim=0pt 0pt {0.33\width} 0pt,clip}{\includegraphics[width=0.6\textwidth]{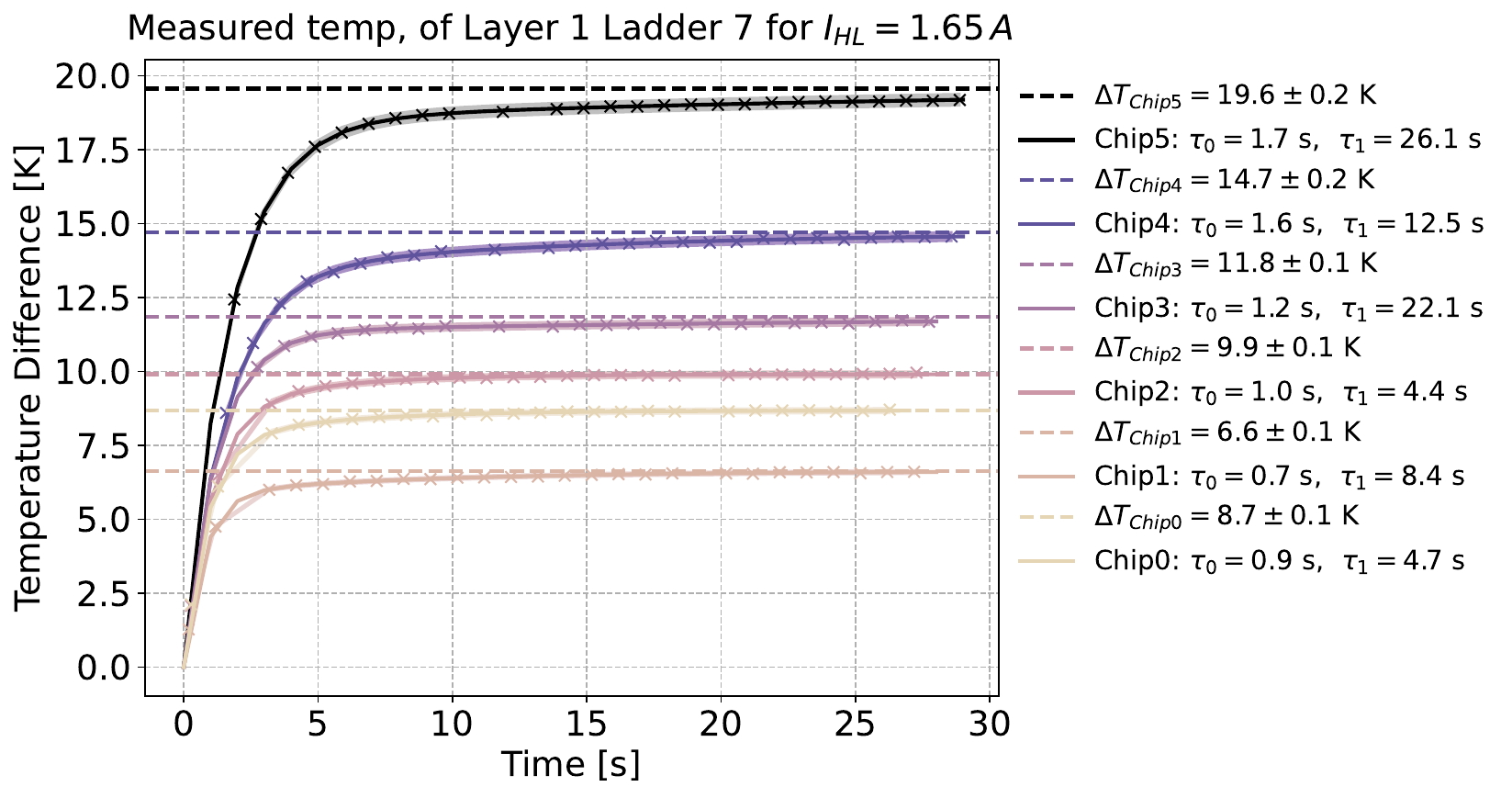}}
\caption{Thermal studies of the innermost two tracking layers of Mu3e. Left: 2-dimensional projection of the chip temperatures of the two layers for a heat load of \qty{350}{\milli\watt/\centi\meter^2}. Gas inlet from left, temperature differences to gas inlet given. Right: Transient temperature behavior of one detector ladder (6 chips) for a heat load of around \qty{215}{\milli\watt/\centi\meter^2}.~\cite{t}\label{fig:cooling}}
\end{figure}

As reported in~\cite{t}, thermal studies were conducted for the inner two tracking layers of Mu3e exploiting a silicon heater mock-up. 
In addition, a detector prototype with a simplified geometry was operated utilizing gaseous helium cooling for the first time.
The mock-up matches the detector in materials and dimensions. 
The silicon heater chips are uniformly heated, and the temperature of each single chip is measured by an on-chip resistive thermometer.
\autoref{fig:cooling} summarizes the main part of these cooling studies.

The maximum chip temperatures reached are around \qty{50}{\kelvin} above the gas inlet temperature for a heat dissipation of \qty{350}{\milli\watt/\centi\meter^2}\footnote{\qty{350}{\milli\watt/\centi\meter^2} is the maximum allowed on-chip heat load for the pixel detectors.}.
With a baseline inlet temperature of \qty{0}{\degreeCelsius}, the chip temperatures are well below the temperature limit of the adhesives used of \qty{+70}{\degreeCelsius}.
The actual heat dissipation expected for \textsc{MuPix11} chips is around \qty{215}{\milli\watt/\centi\meter^2} which reduces the temperature gradient shown in \autoref{fig:cooling} and adds a great safety margin against overheating.
The heat dissipation and chip temperature were found to be linear.
Thus, the temperature profiles from the above measurements can be estimated for lower heat loads .
The chip temperatures stabilize within seconds which ensures a stable operation after start-up, and a stable thermally induced deformation which can be accounted for by track-based alignment. 
Lastly, a detector prototype with PCB-based support, instead of the aluminium-polyimide laminates, equipped with \textsc{MuPix10} chips was successfully operated as the first active pixel detector with gaseous helium cooling as reported in~\cite{intrun}.

\section{MuPix11}

\textsc{MuPix11} is a revision of its predecessor \textsc{MuPix10} \cite{Augustin:2020pkv}, which is the first chip designed to fulfill the specifications of phase I of the Mu3e experiment which includes a hit efficiency of \qty{99}{\percent} and a time resolution < \qty{20}{\nano\second}~\cite{tdr}.
The improvements are the successful operation of the readout at full speed and the custom slow control configuration interface \cite{Augustin:phd}.  

\begin{figure}[tbp]
\centering
\includegraphics[width=.47\textwidth]{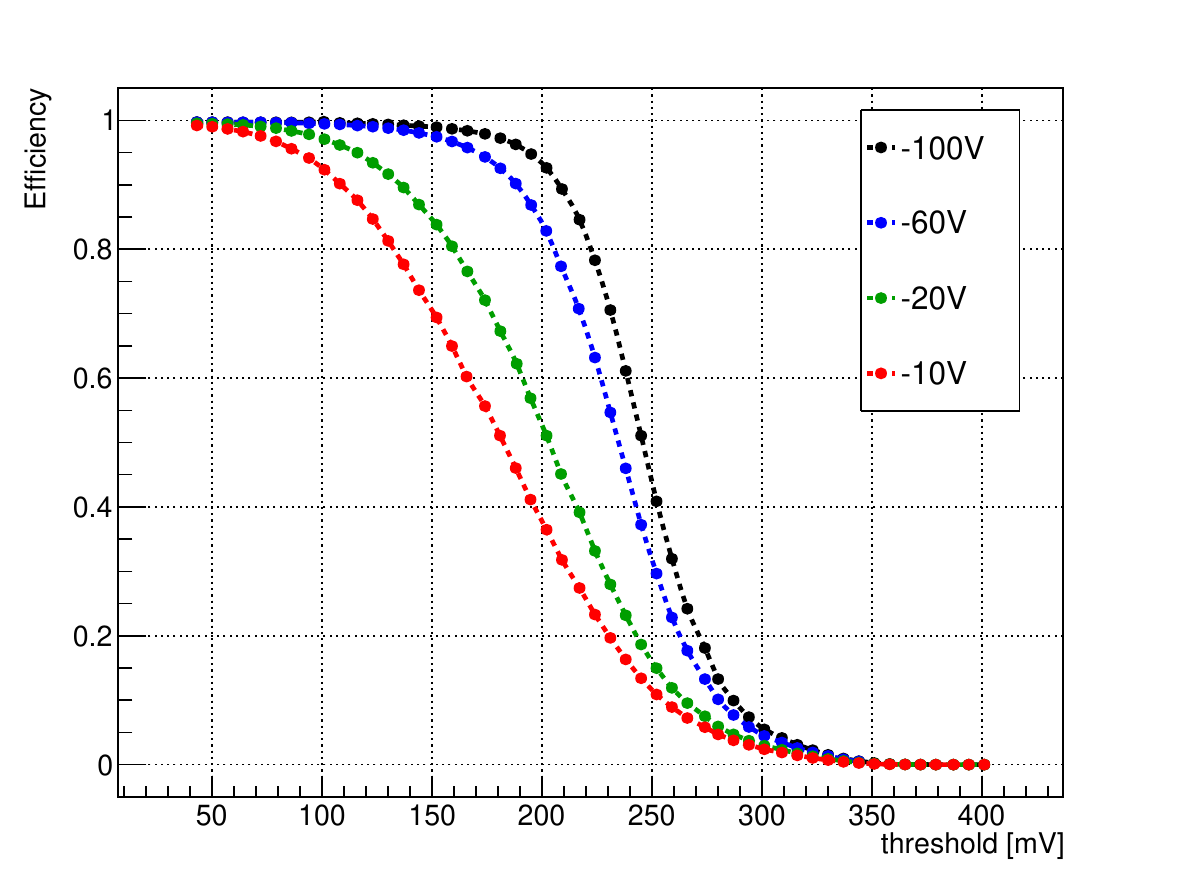}
\qquad
{\includegraphics[width=0.47\textwidth]{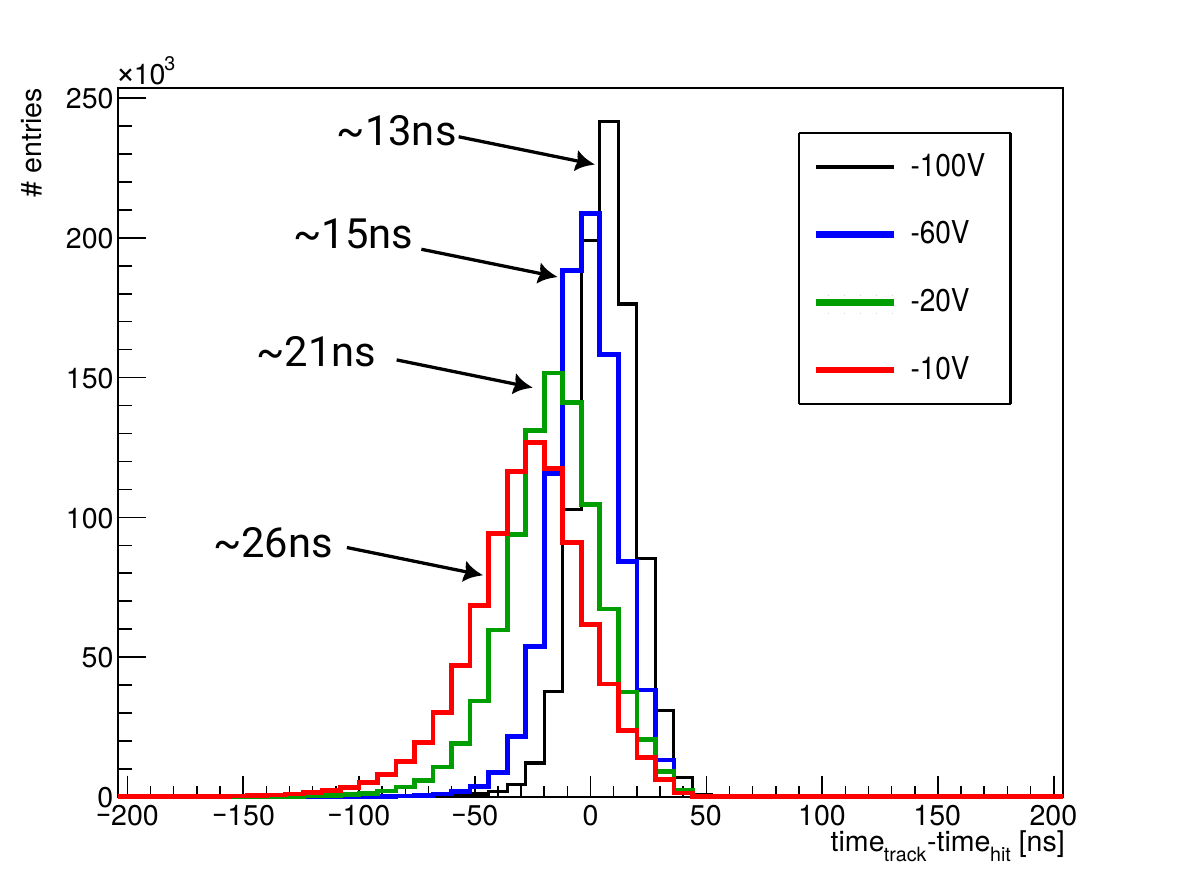}}
\caption{First performance results of a \SI{100}{\micro\meter} thick \textsc{MuPix11} with a substrate resistivity of \SI{370}{\ohm\centi\meter} at different bias voltages and a beam particle momentum of \SI{350}{\mega\electronvolt/c}. Left: Hit detection efficiency as a function of the comparator threshold.  Right: Uncorrected time difference between the track timestamp and its associated hit cluster timestamps for the lowest threshold data point. Indicated is the standard deviation of each corresponding distribution, respectively.}
\label{fig:mp11_results}
\end{figure}

Promising performance results from the commissioning test beam\footnote{ $\pi$M1 beamline at Paul Scherrer Institut: https://www.psi.ch/de/sbl/pim1-beamline} are presented in \autoref{fig:mp11_results}, showing that a \SI{100}{\micro\meter} thick \textsc{MuPix11} meets the efficiency and time resolution requirements of the Mu3e experiment \cite{tdr}. 
At the time of writing, the performance of \SI{50}{\micro\meter} (inner layers) and \SI{70}{\micro\meter} (outer layers) thick \textsc{MuPix11} is under investigation.

\acknowledgments

TTR and DMI thank the HighRR research training group [GRK 2058] for their support. 
We would like to thank the Paul Scherrer Institute for providing high-rate test beams under
excellent conditions.

\end{document}